\documentstyle[12pt,epsfig,a4]{article}
\newcommand{\scite}{~\cite}
\def\beq{\begin{equation}}
\def\eeq{\end{equation}}
\def\beqa{\begin{eqnarray}}
\def\eeqa{\end{eqnarray}}
\def\rd{{\mathrm d}}

\def\pt{p_T}
\def\jpsi{{J\!/\!\psi}}
\def\as{\alpha_s}

\def\bentarrow{\:\raisebox{1.1ex}{\rlap{$\vert$}}\!\rightarrow}
\def\dk#1#2#3{
\begin{array}{r c l}
#1 & \rightarrow & #2 \\
 & & \bentarrow #3
\end{array}
}

\newcommand{\eqr}[1]{~(\ref{#1})}
\begin{document}

\begin{titlepage}

\vspace*{4cm}
\begin{center}
{\Large PHENOMENOLOGY OF ``ONIUM'' PRODUCTION
\footnote{Invited talk at the XXX
Rencontres de Moriond, Les Arcs 1800, March 19-26, 1995. To appear in the
Proceedings.}
}\\
\vspace*{36pt}
{\large Matteo Cacciari
\footnote{Address after May 24th, 1995: DESY, Notkestrasse 85 - 22603
Hamburg - Germany}
} \\
{\sl INFN and Dipartimento di Fisica Nucleare e Teorica,
Universit\`a di Pavia, Italy \\
E-mail: cacciari@pv.infn.it }\\

\vspace{6cm}
\begin{abstract}
The phenomenology of heavy quarkonia production in hadron collisions is
reviewed.
The theoretical predictions are compared to data. Commonly used production
models are shown to fail in explaining all the experimental findings.
The shortcomings of these models are analysed and
possible improvements are discussed.
\end{abstract}

\end{center}

\vfill
\begin{flushleft}
\begin{tabular}{l}
		FNT/T-95/14 \\
		hep-ph/9505332 \\
            May 1995\\
\end{tabular}
\end{flushleft}
\vspace{.5cm}

\end{titlepage}

\section{Introduction}

The production of heavy quarkonium states in high energy processes probes the
very border between
perturbative and non perturbative domains. It provides therefore
quite stringent tests of our understanding of Quantum Chromodynamics (QCD)
and has hence recently
attracted much of the theoretical and experimental interest.

A large amount of data is now available, both from fixed target\scite{mendez}
and from collider\scite{CDF,D0} experiments. In this talk I'll restrict
myself to the
inelastic hadroproduction case, and in particular I'll focus on the collider
results. This because that's where the most serious discrepancies between
theory and experiments have shown up, demanding for a careful reanalysis of the
quarkonium production mechanisms.

The outline of the talk is as follows. I will first describe the general
framework which is believed to encompass quarkonium production, and the way
this
framework has in the past been formalized into production models. The
approximations which had been made will be pointed out,
and the results of theoretical
predictions based on these models will be compared to experimental data and
found to be sometimes in disagreement. Possible reasons for the discrepancies
will then be considered, and a new model which may help reconciling theory with
experiments will be briefly presented and discussed.

\section{Production models}

When considering any bound state object production two key questions have to be
answered:
\begin{enumerate}
\item What is it made of?
\item How are its components produced?
\end{enumerate}

As for the first question, there is evidence that heavy quarkonia are (mainly)
composed of a heavy quark-antiquark pair, i.e. $c\bar c$ for charmonium (the
$\jpsi$ family) and $b\bar b$ for the $\Upsilon$ family.

As for the second questions, there seems to be general agreement that
QCD is {\sl the} theory of strong interactions. It is therefore
within its context that the production of the heavy quark and antiquark has to
be described.

These two questions and the answers given above constitute the general
framework
to quarkonium phenomenology. At this points the different models we can think
of
do diverge and provide different descriptions and hence different results. I'll
now describe two models which have been put forward some years ago and
widely used till today: Fritzsch's Color Evaporation Model\scite{cem} (CEM)
and Berger and
Jones' Color Singlet Model\scite{csm} (CSM).

\subsection{The Color Evaporation Model}

This model makes
use of the  parton-hadron duality hypothesis to relate the
charmonium cross section to the quark-antiquark production cross section. In
particular, it assumes that all the non perturbative effects that lead to the
bound state formation cancel when considering inclusive final states, and
therefore writes the production cross section for the quarkonium state $H$ as:
\beq
\sigma[H] = P_H\int_{4m_Q^2}^{4m_M^2}\rd s\;\sigma[Q\bar Q]
\label{cem}
\eeq
On the right hand side we have the QCD cross section for producing a
$Q\bar Q$ pair, $\sigma[Q\bar Q]$, integrated over the invariant mass range up
to the mass $m_M$ of the lowest lying heavy-light meson (for instance the
$D$ meson in
the charm case). The factor $P_H$ has to be determined phenomenologically and
provides the only differentiation among the various quarkonium states. While
being therefore quantitatively not very predictive,
this model has its central feature in that the differential
distributions (like energy, $x_F$, or $\pt$ dependencies) should be equal for
all the quarkonium states considered. It is thus mainly on this ground that its
degree of validity will have to be assessed. For a review of this model and its
achievements see for instance ref.~\cite{schuler}.

\subsection{The Color Singlet Model}

This model tackles the problem of quarkonium
production from an opposite point of view with respect to the Color Evaporation
Model. It tries to reach the highest possible predictivity at the price of
making stronger approximations.

Its central assumption is that the $Q\bar Q$ pair that will form the bound
state
is produced by the hard (short distance) QCD interaction with the correct spin
and angular quantum numbers for the quarkonium considered, in a color singlet
(and therefore observable) configuration and with zero or small relative
momentum:
\beq
\sigma[n\,^{2S+1}L_J] = P_{nL}\,\sigma[Q\bar Q(n\,^{2S+1}L_J,\underline{1})]
\label{csm}
\eeq
The non perturbative effects are
assumed to factorize into a single parameter $P_{nL}$,
which can be related to the
wave function of the bound state (or its derivative, for $P$ states) evaluated
at the origin. It can be calculated within potential
models (see for example ref.\cite{quigg}) or extracted from experimental data
of
quarkonium decay widths, which can be calculated within the same model and
yield
expressions analogous to\eqr{csm}.

The difference with the previous model can easily be appreciated: the structure
of the quarkonium (``What is it made of?'') is now defined much more precisely,
and this allows detailed quantitative
calculations for each different state $H = n\,^{2S+1}L_J$. On the other hand,
to assume that a quarkonium is
well described by a $Q\bar Q$ pair in a color singlet configuration may be
too a crude approximation.

\section{Phenomenology}

When we make use of the CSM to obtain phenomenological predictions for
quarkonia
production we have to live with a double approximation.

The first one is
intrinsic in the definition of a quarkonium as a color singlet $Q\bar Q$ pair
with zero relative momentum:
this amounts to neglect relativistic corrections, i.e. to work at lowest order
in $v^2$, $v$ being the heavy quark speed in the quarkonium center of mass
frame.

The second is the perturbative order at which the cross section for producing
the heavy quark-antiquark pair is evaluated. The lowest order in $\as$ was
usually considered to represent the dominant term in the expansion.

When calculating large $\pt$ production rates for quarkonia at the Tevatron,
i.e. in $p\bar p$ collisions at $\sqrt{s} = 1800$~GeV, within the above
approximations we immediately find {\sl huge} discrepancies between theory and
experiment. The $\jpsi$ and $\psi'$ rates are found to be underestimated by one
or two orders of magnitudes.

We may at this point ask ourselves if are we operating within a wrong model or
if we are just neglecting important contributions when making the
aforementioned approximations. I'll pursue the second point of view and try to
present the CSM as being the lowest order approximation of a wider model. It
must therefore be supplemented by higher order terms, \underline{both in $\as$
and
in $v^2$}, in order
to produce reliable predictions.

Following the historical development I start considering higher order terms in
$\as$. In 1993 Braaten and Yuan\scite{bygluon} pointed out that the process
\beq
\dk{gg}{\;gg}{\jpsi\;gg}
\label{gfrag}
\eeq
while being of higher order ($\as^5$ vs. $\as^3$) with respect to the lowest
order one,
\beq
gg\to\jpsi\;g
\label{gfus}
\eeq
becomes however dominant at large $\pt$ (say $\pt >$ 5--6 GeV). The technical
reason for this is that the process\eqr{gfrag} is not suppressed by a form
factor $(m/\pt)^2$, $m$ being the heavy quark mass, which is instead present in
the cross section of\eqr{gfus}. The physical reason for the existence of this
form factor is that it is difficult to produce the heavy quark and
antiquark in a small spatial region, of size $1/\pt$, and still keep their
relative momentum close to zero so that they can form a bound state.
This shortcoming is absent in the process\eqr{gfrag}, as the
gluon that fragments to the $\jpsi$ can have invariant mass of order $m\ll\pt$
and thus make the heavy quark-antiquark formation a longer distance process.
Moreover, they could also show that in the limit of the gluon energy being much
larger than its invariant mass the cross section could be approximated by
the convolution of a kernel cross section for producing on-shell gluons with a
{\sl fragmentation function} for the gluon going to the $\jpsi$.

The
fragmentation functions for a gluon or a heavy quark to go into any quarkonium
state can be calculated\scite{bygluon,ds} in perturbative QCD
(with the exception of the usual
non-perturbative parameters related to the bound state formation) and the
cross section  is given by
\beq
{{\rd\sigma[H(\pt)]}\over{\rd\pt}} =
\int{{\rd\hat\sigma[i(\pt/z),\mu]}\over{\rd\pt}}D_i^H(z,\mu)
\eeq
In this equation $\hat\sigma$ is the kernel cross section for producing the
parton $i$, and $D_i^H$ is the fragmentation function of $i$ to the quarkonium
$H$. $\mu$ is the factorization scale, to be taken of the order of $\pt$.

\begin{figure}[t]
\vspace{-.3cm}
\begin{center}
\begin{minipage}{7cm}
\end{minipage}
\begin{minipage}{7cm}
\end{minipage}
\parbox{12cm}{
\caption{\label{cdf-prel}\small Theoretical prediction for the
 $\jpsi$ (a) and $\psi'$ (b) cross sections compared to CDF experimental data.}
}
\end{center}
\end{figure}

Phenomenological predictions based on this formula\scite{frags} have promptly
shown that the CDF data\scite{CDF} on $\jpsi$ production could now
be explained within a
factor of two to five, depending on the input parameters, (see
fig.~\ref{cdf-prel}a)
while predictions for
$\psi'$ productions where still falling a factor of 30 below the data (see
fig.~\ref{cdf-prel}b).

The solution to this discrepancy, which has been named the ``CDF Anomaly'', may
have been found within the context I'm going to describe in the next Section.

\section{The Factorization Approach}

A rigorous frame for treating quarkonium production and decays has been
recently
developed by Bodwin, Braaten and Lepage\scite{bbl}.
Their so-called ``factorization model''
expresses the cross section for quarkonium production as a sum of terms each of
which contains a short-distance perturbative factor and a long-distance non
perturbative matrix element:
\beq
\sigma[H] = \sum_n {{F_n(\Lambda)}\over{m^{\delta_n-4}}} \langle 0|{\cal
O}^H_n(\Lambda)|0\rangle
\label{x-fact}
\eeq
$F_n$ are short-distance coefficients which can be calculated in
perturbative QCD by expanding in powers of
$\as$.  $\Lambda$ is a
scale which separates short and long distance effects.  The cross section
is however independent of $\Lambda$ as its effect is compensated by the
$\Lambda$-dependence of the non-perturbative matrix elements. $\delta_n$ are
related to the dimension of the operator ${\cal O}^H_n$.   Finally, the
matrix elements $\langle 0|{\cal O}^H_n(\Lambda)|0\rangle$ can be defined
rigorously in Non Relativistic QCD. They absorb the non perturbative
features of the process and can either be extracted from data or calculated on
a
lattice.

The main feature of this model, and the main difference with respect to the
CSM,
is that - pretty much like the Color Evaporation Model but in a much more
sophisticated way - it takes into account the full Fock space structure of
the quarkonium
state. The latter is no more assumed to be represented by a color singlet
$Q\bar Q$ pair with the correct quantum numbers, but rather by an infinite
series of terms:
\beqa
|H=n\,^{2S+1}L_J\rangle&=&O(1)|Q\bar Q(n\,^{2S+1}L_J),\underline{1}
\rangle\nonumber\\
&+& O(v)|Q\bar Q(n\,^{2S+1}(L\pm 1)_{J'},\underline{8}) g\rangle\nonumber\\
&+& O(v^2)|Q\bar Q(n\,^{2S+1}L_J,\underline{8})gg\rangle + ...\nonumber\\
&+& ...
\eeqa
The CSM can be recovered by taking the lowest order term in eq.\eqr{x-fact}.

Higher order components are suppressed by powers of $v$, but can become
important if their associated short-distance coefficient $F_n$ in
eq.\eqr{x-fact} is large. Braaten and Fleming\scite{bf} have shown
that $\psi$ production
via gluon fragmentation through a color octet $^3S_1$ pair, being its short
distance coefficient of order $\as$ only, can easily overwhelm ordinary gluon
fragmentation to a color singlet pair (of order $\as^3$) and may thus explain
the $\psi'$ abundance that CDF observes.

\section{Conclusions}

In this talk I have presented evidence that lowest order Color Singlet Model
results are
incapable of explaining CDF data of quarkonium production. Higher order terms
both in $\as$ and in $v^2$ are important and must be included to ensure a
consistent and satisfactory description of quarkonia production.

The interplay between higher orders in $\as$ and in $v^2$ is far from being
trivial, and quarkonia production has to be regarded as a
two-parameter problem.  A consistent framework for treating the double
expansion can be found
in the factorization model developed by Bodwin, Braaten and Lepage.

\section*{Acknowledgements}

It is a pleasure to thank E.L. Berger for the invitation and the Organizing
Committee for the pleasant staying during the Conference. I also thank M. Greco
for the numerous and fruitful discussions about this subject.

\end{document}